\documentclass[conference]{IEEEtran}

\usepackage{cite}
\usepackage{setspace}
\usepackage{graphicx}

\usepackage[cmex10]{amsmath}
\usepackage{amssymb}

\usepackage{array}
\usepackage{booktabs}
\usepackage{threeparttable}
\usepackage[update,prepend]{epstopdf}
\usepackage{multirow}
\usepackage{amsmath}

\usepackage{epsfig}
\usepackage{indentfirst}

\usepackage{subfigure}
\usepackage{float}
\usepackage{multirow}

\begin{document}

\title{\bfseries \LARGE
On Timing Synchronization for Quantity-based Modulation in Additive Inverse Gaussian Channel with Drift}

\author{\IEEEauthorblockN{Bo-Kai Hsu$^\dagger$, Chia-Han Lee$^\dagger$, and Ping-Cheng Yeh$^\dagger$}\\
\IEEEauthorblockA{$^\dagger$Communications and Networks Science (CoWorkS) Laboratory\\
$^\dagger$Department of Electrical Engineering and
Graduate Institute of Communication Engineering\\
National Taiwan University\\
$^\ddagger$Research Center for Information Technology Innovation\\
Academia Sinica, Taiwan
}}

\maketitle

\begin{abstract}
In Diffusion-based Molecular Communications, the channel between Transmitter Nano-machine (TN) and Receiver Nano-machine (RN) can be modeled by Additive Inverse Gaussian Channel, that is the first hitting time of messenger molecule released from TN and captured by RN follows Inverse Gaussian distribution. In this channel, a quantity-based modulation embedding message on the different quantity levels of messenger molecules relies on a time-slotted system between TN and RN. Accordingly, their clocks need to synchronize with each other. In this paper, we discuss the approaches to make RN estimate its timing offset between TN efficiently by the arrival times of molecules. We propose many methods such as Maximum Likelihood Estimation (MLE), Unbiased Linear Estimation (ULE), Iterative ULE, and Decision Feedback (DF). The numerical results shows the comparison of them. We evaluate these methods by not only the Mean Square Error, but also the computational complexity.

\end{abstract}


\section{Introduction}

In recent years, nanotechnology is developing rapidly. It become possible to manufacture a device in scale of molecules, nano-machine. Because of the size, the computational capability and the memory of a nanomachine will be limited. Accordingly, to reach a functional system, we need to form a nanonetwork by communicating and cooperating between nanomachines. In a nanoscale network, one of promising end-to-end communication approach is molecular communication, which propagates information by transmitting and receiving messenger molecules. 

In Diffusion-based Molecular Communications (DMC), molecules diffuse across a fluid medium from regions of high to low concentration. This process can be modeled by Fick's laws of diffusion and Brownian motion process\cite{2012Sha}. In this field, many papers design a good modulation and detection to improve the quality of molecular communication. For example, the paper \cite{lin_2012_signal} considers a time-slotted diffusion-based molecular communication with information embedded in different quantity levels. However, most literature assume perfect synchronization between the transmitter and the receiver. In reality, how to form a time-slotted system in DMC is still a problem. To solve this problem, we analyze the timing synchronization problem in this paper.

Generally, transmitter converts information bit stream into a sequence of symbols. Then, transmitter assigns a period of time called \emph{symbol duration} to transmit each symbol. However, with non-synchronous clocks of transmitter and receiver, A timing offset which is constant but unknown for receiver exists between these two clocks. Accordingly, receiver may not identify the beginning of each symbol duration. This is the problem of \emph{timing synchronization}  or \emph{symbol synchronization} \cite{2009Sha}.

For timing synchronization problem in concentration-based molecular communication, the first blind synchronization algorithm has been proposed in \cite{2013Sha}. This paper use the concentration single measured by receiver to efficiently estimate the propagation delay of transmission. But our system model is different from this paper's. The situation in our transmission is releasing a few number of molecules, which is less than the level to form a concentration single.  Then, when molecules arrive to the position of receiver, they will be captured one by one and receiver can measure the arrival times of molecules. These two types of system model of communication in DMC has been studied in literature \cite{ lin_2012_signal , ecford_2007_nanoscale }.

The key contributions of this paper are listed as below. First, we compare the Mean Square Error (MSE) and computational complexity of three methods in \emph{training-based synchronization}, Maximum Likelihood Estimation (MLE) ,Unbiased Linear Estimation (ULE), and Iterative ULE. Among them, we proposed the best one, Iterative ULE, with lower complexity and its MSE reaches almost the same efficient level. On the other hand, in \emph{blind synchronization}, we analyze the theoretical MSE of ULE for the first arrival time and Decision Feedback (DF) method to give a sufficient condition when the latter improves the former.

The following structure of our paper begins with system model and problem formulation in Sec.~\ref{sec:model}. We will explain what Additive Inverse Gaussian Channel (AIGC) and quantity-based modulation are in DMC. Then, in Sec.~\ref{sec:training} and Sec.~\ref{sec:blind}, we discuss the timing offset estimation in training-based and blind synchronization, respectively. The next Sec.~\ref{sec:numerical} shows the simulation MSE and theoretical curves of all proposed methods. Finally, the conclusion and future work are discussed in Sec.~\ref{sec:conclusion}.

\section{System Model And Problem Formulation} \label{sec:model}

We consider an end-to-end communication in a volume with fluid medium. The transmitter is a nano-machine, and so is the receiver. We call them Transmitter Nano-machine (TN) and Receiver Nano-machine (RN). They communicate with each other by releasing and capturing molecules. The channel between them is molecular diffusion based on random walk to propagate information message. Based on \cite{chhikara_1989_inverse}, a one-dimensional molecular diffusion can be described by Brownian motion and the first hitting time to a specific position follows the Inverse Gaussian distribution.

We define the spatial location by a one-dimensional extent with the origin as TN's position. Let $d > 0$ denote the position of RN apart from TN. When a molecule is released from TN at time $x$, it will act as one-dimensional Brownian motion with positive drift velocity $v > 0$ in the medium. Because of positive drift, in a sufficiently long period of time, this molecule must arrive in the RN's position and be captured by RN. However, the arrival time of this molecule is random.  Based on \cite{chhikara_1989_inverse}, we describe the arrival time as $x + T$, where the additive first hitting time $T$ is the random variable with Inverse Gaussian distribution.
\begin{equation} \label{eq:def_Ik}
f_T(t) =
\begin{cases}
\sqrt{\frac{\lambda}{2 \pi t^3}}\exp\{-\frac{\lambda(t-\mu)^2}{2 \mu^2 t}\} & \text{if } t>0 , \\
0 & \text{if } t \leq 0 .
\end{cases}
\end{equation}
with parameters as follows:
\begin{IEEEeqnarray}{rCl} 
\mu = \dfrac{d}{v}\text{\ ,\ } \lambda = \dfrac{d^2}{2 D} \text{\ and,\ }
D = \dfrac{k_BT_a}{6\pi \eta r}, \nonumber
\end{IEEEeqnarray}
where $k_B$ is the Boltzmann constant, $T_a$ is the absolute temperature, $\eta$ is the viscosity of the fluid medium, and $r$ is the radius of molecule. This channel is known as Additive Inverse Gaussian Channel (AIGC).

In AIGC, an optimal detection in quantity-based modulation is proposed in\cite{lin_2012_signal} to improve quality of communication. Because this modulation and detection rely on a time-slotted system, we consider the synchronization problem in quantity-based modulation. That is, message is conveyed by a sequence of symbols in consecutive symbol duration $T_s$ and TN assign the quantity of molecules as different symbols. All molecules are in the same type, so RN cannot distinguish them. In binary case, TN release $L_0$ and $L_1$ unlabeled molecules representing a binary zero and a binary one, respectively. Then, RN will accumulate the quantity of arrival molecules during symbol duration $T_s$ and demodulate this symbol by some criteria.
\begin{figure}[!h]
\centering
\includegraphics[scale=0.5]{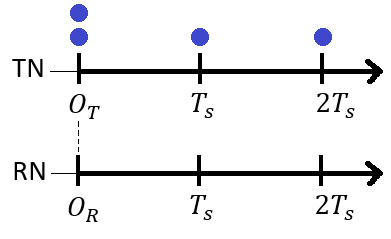}
\caption{Time-slotted system in quantity-based modulation under perfect synchronization.}\label{fig:tims-slotted}
\end{figure}
\begin{figure}[!h]
\centering
\includegraphics[scale=0.4]{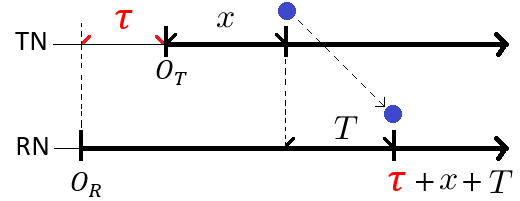}
\caption{Phenomenon of timing offset.}\label{TxRx}
\end{figure}

Fig.~\ref{fig:tims-slotted} shows time-slotted system in quantity-based modulation under perfect synchronization. However, in the beginning of communication, TN and RN may have non-synchronous clocks with each other. We denote the starting point of TN's and RN's clock by $O_T$ and $O_R$ respectively. As shown in Fig.~\ref{TxRx}, a timing offset $\tau:=O_T-O_R$ is defined as subtraction $O_R$ from $O_T$, which may be negative, to represent the non-synchronous phenomenon. Consequently, when RN receive a molecule, the arrival time measured by RN includes not only the random delay $T$, but also the timing offset $\tau$ which is constant but unknown for RN. The problem is how to efficiently estimate the timing offset $\tau$ by the sequence of arrival time measured by RN to reach synchronization with TN.
\begin{figure}[!h]
\centering
\includegraphics[scale=0.4]{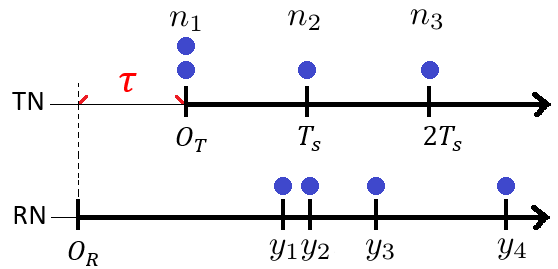}
\caption{Timing offset estimation for quentity-based modulation.}\label{fig:estimation}
\end{figure}

As shown in Fig.~\ref{fig:estimation}, TN releases $n_1, n_2, ..., n_K$ molecules in the beginning of consecutive $K$ symbols. In binary case, $n_k \in \{L_0, L_1\}$. What RN observe is a sequence of arrival times of molecules denoted by $y_1, y_2, ..., y_{n_1+n_2+...+n_K}$ according to the order of timing. Based on the observation $\mathbf{y}=[y_1, y_2, ..., y_N]$, where $N=n_1+n_2+...+n_K$, we want to design an efficient estimator $\hat{\tau}(\mathbf{y})$ to make the MSE defined by $E[(\tau - \hat{\tau}(\mathbf{y}))^2]$ as small as possible, where the function $E[.]$ denotes the expectation of some random variable. 
\begin{figure}[!h]
\centering
\includegraphics[scale=0.3]{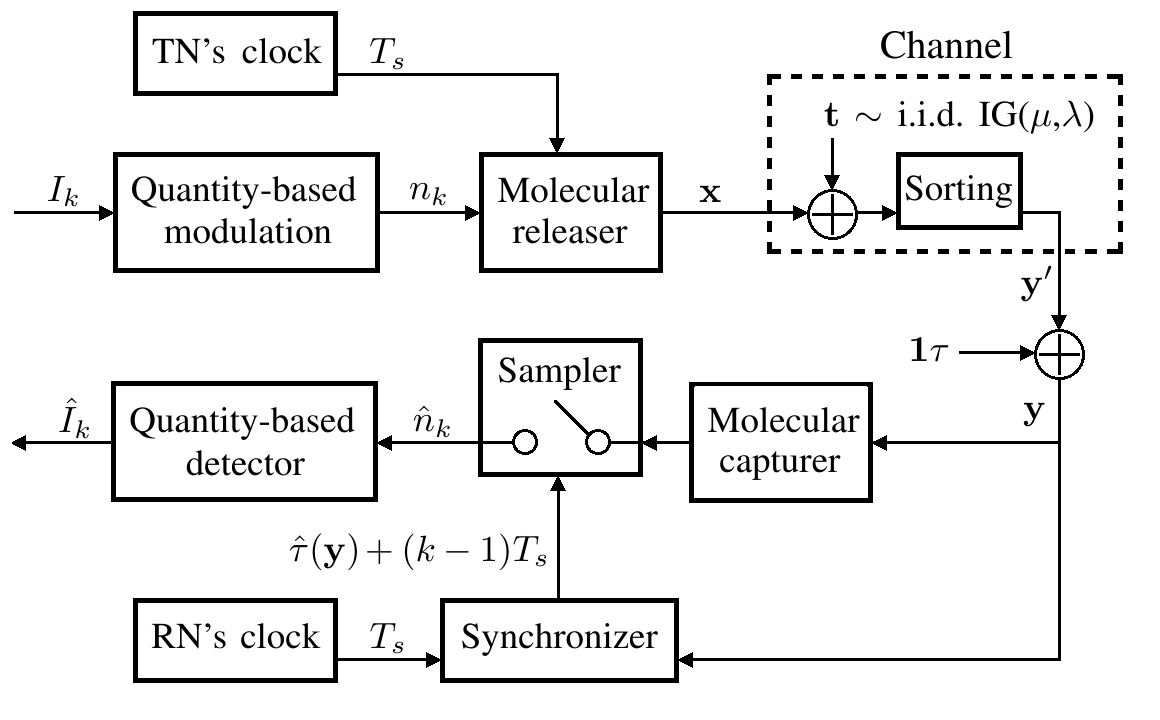}
\caption{System diagram of timing offset estimation for quantity-based modulation.}\label{fig:diagram_time}
\end{figure}

Fig.~\ref{fig:diagram_time} shows an overall system diagram of timing offset estimation for quantity-based modulation. In this system, TN modulates information bit stream $I_k$ into the sequence of molecular amounts for each symbol $\{n_k|1 \leq k \leq K\}$. Then, TN releases molecules based on TN's clock. We denote the molecular releasing time sequence as $\mathbf{x}=[x_1, x_2, ..., x_N]$, where $N=n_1+n_2+...+n_K$. For quantity-based modulation, $x_i = (j-1)T_s$ if $n_1+...+n_{j-1}<i\leq n_1+...+n_j$ for $i \in \{1,2,...N\}$. For example, if $K=3$ and $n_1=2,n_2=1,n_3=1$, as shown in Fig. \ref{fig:tims-slotted}, we have $\mathbf{x}=[0,0,T_s,2T_s]$. After passing through diffusion channel, $\mathbf{x}$ becomes $\mathbf{y}'$. Because of the non-synchronous clock between TN and RN, the actual arriving time measured by RN $\mathbf{y}$ includes the unknown timing offset $\tau$. Accordingly, we want to design an efficient estimator $\tau(\mathbf{y})$ by the observation $\mathbf{y}$ to make the sampling time $\tau(\mathbf{y}) + (k-1)T_s$ synchronize with TN. 

In the next two sections, we will discuss on two types of synchronization: training-based and blind synchronization.

\section{Training-based synchronization }\label{sec:training}

In training-based synchronization, there is a training phase to synchronize between TN and RN before they transmit and receive information message. In training phase, TN transmit a pilot signal which RN have already known, so $\mathbf{x}$ is constant for RN. The following we propose two methods to estimate $\tau$ under the assumption that RN knows $\mathbf{x}$.


\subsection{Maximum Likelihood Estimation}

Based on \cite{ecford_2007_nanoscale}, when $\tau = 0$, that is perfect synchronization, the joint probability density function (pdf) of observation denoted by $\mathbf{y'}$ given releasing time sequence $\mathbf{x}$ has been derived as below:
\begin{IEEEeqnarray}{rCl} 
f(\mathbf{y'}|\mathbf{x}) 
&=& \displaystyle\sum_{ \mathbf{u} \in \mathbf{P}(\mathbf{y'})} f(\mathbf{u}|\mathbf{x}) \nonumber \\
&=&
\begin{cases}
\displaystyle\sum_{\mathbf{u} \in \mathbf{P}(\mathbf{y'})} \prod_{i=1}^{N} f_T(u_i-x_i), &\text{if } \mathbf{y'}=\text{sort}(\mathbf{y'}); \\
0, &\text{otherwise}.
\end{cases}
\end{IEEEeqnarray}
where $\mathbf{P}(\mathbf{y'})$ is the set of all possible permutation of $\mathbf{y'}$ and the function $\text{sort}(\mathbf{y'})$ sorts $\mathbf{y'}$ according to ascending order. 

In our work, because of the non-synchronous phenomenon, the observation of RN is $\mathbf{y}=\mathbf{y'}+\tau\mathbf{1}_N$, where $\mathbf{1}_N$ is the $1 \times N$ vector with all value are equal to $1$. Therefore, we can derive likelihood function as below:
\begin{IEEEeqnarray}{rCl} 
f(\mathbf{y}|\mathbf{x},\tau)=
\begin{cases}
\displaystyle\sum_{\mathbf{u} \in \mathbf{P}(\mathbf{y})} \prod_{i=1}^{N} f_T(u_i-x_i-\tau), &\text{if} \mathbf{y}=\text{sort}(\mathbf{y}); \nonumber \\
0, &\text{otherwise}.  \nonumber \\
\end{cases} 
\nonumber \\
\end{IEEEeqnarray}
Then, we can apply MLE as our first estimator $\hat{\tau}_{\textnormal{MLE}}$:
\begin{IEEEeqnarray}{rCl} 
\hat{\tau}_{\textnormal{MLE}} &:=& \arg\max_{\tau} f(\mathbf{y}|\mathbf{x},\tau).
\end{IEEEeqnarray}
However, the time complexity grows rapidly by factorial on $N$ because of the permutation of $\mathbf{y}$. In reality, it beyond the computational capability of nano-machine.

\subsection{Unbiased Linear Estimation}

By considering the complexity issue, we try to apply Linear Estimation (LE) on the first $n$ observations, $\mathbf{y}_n = [y_1, y_2, ..., y_n]$, where $n \leq N$. Assume $\hat{\tau}_{\textnormal{ULE}}:=\mathbf{a}_n\mathbf{y}_n^{\top}+b$ for some constant $\mathbf{a}_n = [a_1, a_2, ...,a_n]$ and $b$ such that MSE is minimal.
%

We have $\mathbf{y}_n = \mathbf{y'}_n + \tau\mathbf{1}_n$, where $\mathbf{y'}_n$ is the first $n$ components of the random vector $\mathbf{y'}$. By derivation above, we have the joint pdf of $\mathbf{y'}_n$ given $\mathbf{x}$, so we can derived the mean and covariance of $\mathbf{y'}_n$, which are used to derive the coefficient $\mathbf{a}_n$ and $b$.

Because the parameter $\tau$ is constant but unknown, we set the constraint on $\mathbf{a}_n\mathbf{1}_n^{\top} = 1$ to eliminate $\tau$ as below:
\begin{IEEEeqnarray}{rCl} 
E[{(\tau - \hat{\tau}_{\textnormal{ULE}})}^2]
&=& E[{(\tau - \mathbf{a}_n\mathbf{y}_n^{\top}-b)}^2] \nonumber \\
&=& E[{(\tau -\mathbf{a}_n\mathbf{1}_n^{\top}\tau - \mathbf{a}_n\mathbf{y'}_n^{\top}-b)}^2]
\nonumber \\
&=& E[{(\mathbf{a}_n\mathbf{y'}_n^{\top}+b)}^2].
\end{IEEEeqnarray}
According to \cite{Candan_2011_Nonrandom}, applying the solution in Non-random parameter estimation, we can make MSE reach the minimal value when 
\begin{IEEEeqnarray}{rCl} 
\label{eq:a_n and b}
\mathbf{a}_n &=& \mathbf{1}_n\mathbf{C}_{\mathbf{y'}_n}^{-1}\{\mathbf{1}_n\mathbf{C}_{\mathbf{y'}_n}^{-1}\mathbf{1}_n^{\top}\}^{-1} \text{ and}
\nonumber  \\
b&=&-E[\mathbf{a}_n\mathbf{y'}_n^{\top}] \text{,}
\end{IEEEeqnarray}
where $\mathbf{C}_{\mathbf{y'}_n}$ is the covariance matrix of the random vector $\mathbf{y'}_n$. 

Besides eliminating unknown $\tau$, setting $\mathbf{a}_n\mathbf{1}_n^{\top} = 1$ make the unbiased property possible, so we actually apply LE under the unbiased constraint, which called Unbiased Linear Estimation (ULE) by us.
\begin{IEEEeqnarray}{rCl} 
\label{eq:unbiased}
E[\hat{\tau}_{\textnormal{ULE}}] 
&=& E[\mathbf{a}_n\mathbf{y}_n^{\top}+b] \nonumber \\
&=& E[\mathbf{a}_n\mathbf{1}_n^{\top}\tau + \mathbf{a}_n\mathbf{y'}_n^{\top} + b] \nonumber \\ 
&=& E[\tau] + E[\mathbf{a}_n\mathbf{y'}_n^{\top}] + b = \tau.
\end{IEEEeqnarray}
The last step in \eqref{eq:unbiased} follows by $b=-E[\mathbf{a}_n\mathbf{y'}_n^{\top}]$ in \eqref{eq:a_n and b} and the unknown $\tau$ is constant.

\begin{figure}[h]
	\centering
	\includegraphics[scale=0.4]
{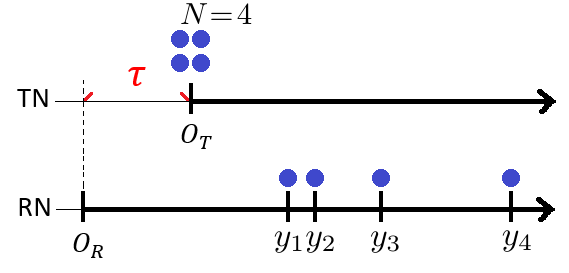}
	\caption{An example for ULE when $\mathbf{x}=\mathbf{0}_4$.}
	\label{train_syn}
\end{figure}

Let's give an example when $n=2$, $K=1$, and $\mathbf{x}=\mathbf{0}_{n_1}$ as a zero vector with length $n_1$, where $n_1=N \geq n=2$. As shown in Fig. \ref{train_syn}, if TN release $n_1$ molecules at the beginning of communication. In this case, the random vector $\mathbf{y'}_n = [T_{(1)}, T_{(2)},...,T_{(n)}]$, where $T_{(i)}$ is the $i$-th order statistic of $n_1$ independent and identical distribution (iid) random variables with generic Inverse Gaussian distribution $f_T(t)$.

Applying the solution of ULE in this case, we get the linear estimator as below:
\begin{IEEEeqnarray}{rCl} 
\hat{\tau}_{\textnormal{ULE}} &:=& \mathbf{a}_2\mathbf{y}_2^{\top}+b \nonumber \\
&=& a_1y_1+a_2y_2-E[a_1T_{(1)}+a_2T_{(2)}] \nonumber \\ 
&=& a_1(y_1-\mu_1)+(1-a_1)(y_2-\mu_2),
\end{IEEEeqnarray}
where $\mu_1 = E[T_{(1)}]$, $\mu_2 = E[T_{(2)}]$, and 
\begin{IEEEeqnarray}{rCl} 
a_1 = \frac{Var[T_{(2)}]-Cov[T_{(1)},T_{(2)}]}{Var[T_{(1)}-T_{(2)}]}, 
\end{IEEEeqnarray}
where the function $Var[.]$ and $Cov[.,.]$ denotes, respectively, the variance of some random variable and the covariance of two random variables.
With this estimator, the theoretical MSE can be derived as below:
\begin{IEEEeqnarray}{rCl} 
E[{(\tau - \hat{\tau}_{\textnormal{ULE}})}^2] &=& Var[\mathbf{a}_2\mathbf{y}_2^{\top}] \nonumber \\
&=& \frac{Var[T_{(1)}]Var[T_{(2)}]-Cov[T_{(1)},T_{(2)}]^2}{Var[T_{(1)}-T_{(2)}]}. \nonumber \\
\end{IEEEeqnarray}
%
%

\subsection{Iterative ULE Per Symbol}

In the above, we proposed ULE, which reduces the complexity of MLE, in training-based synchronization. The ULE only needs linear time computation on $n$ after we have the weighted value $\mathbf{a}_n$. However, in the algorithm of ULE, the complexity of computing the weighted value $\mathbf{a}_n$ beforehand still grows rapidly by factorial on $n$, which is incomputable when $n$ is large. To simplify the computation of $\mathbf{a}_n$ and make linear estimation for large $n$ possible, we rewrite the algorithm to iterative form, which called Iterative ULE (IULE) by us.

First of all, let's recall the linear combination in ULE for the first $n$ arrival time $\mathbf{y}_n$ in \eqref{eq:linear combination}. We denote the estimator of ULE for $n$ as $\hat{\tau}_n$ and introduce a constant vector $\mathbf{u}_n$ to represent the expected value of the random vector $\mathbf{y'}_n$. This way, it is clear that $\mathbf{a}_n$ is the weighted value of $n$ estimators $y_i - u_i$ for $i=1,2,...,n$.
\begin{IEEEeqnarray}{rCl} 
\label{eq:linear combination}
\hat{\tau}_n &:=& \mathbf{a}_n\mathbf{y}_n^{\top}+b 
= \mathbf{a}_n(\mathbf{y}_n-\mathbf{u}_n)^{\top},
\end{IEEEeqnarray}
where $\mathbf{u}_n=E[\mathbf{y'}_n]$.

Assume TN transmits total $K$ symbols with the same quantity of molecules $n_1$. That is ,the training sequence $\{n_k|1 \leq k \leq K\}$ is a constant sequence. When RN receives $k$ symbols, we can apply ULE for $n=k n_1$. When RN receives $k+1$ symbols, we can apply ULE for $n=(k+1)n_1$. We compare these two estimators to derive the iterative form of IULE.
\begin{IEEEeqnarray}{rCl} 
\label{eq:k_and_k+1}
\hat{\tau}_{k n_1}
&:=& \mathbf{a}_{k n_1}(\mathbf{y}_{k n_1}-\mathbf{u}_{k n_1})^{\top}, \nonumber \\
\hat{\tau}_{(k+1)n_1}
&:=& \mathbf{a}_{(k+1)n_1}(\mathbf{y}_{(k+1)n_1}-\mathbf{u}_{(k+1)n_1})^{\top},
\end{IEEEeqnarray}

In \eqref{eq:k_and_k+1}, the first $k n_1$ components of $\mathbf{y}_{k n_1}$ and $\mathbf{y}_{(k+1)n_1}$ are the same, so are $\mathbf{u}_{k n_1}$ and $\mathbf{u}_{(k+1)n_1}$. Accordingly, we denote $\mathbf{y}_{\textnormal{new}}$ and $\mathbf{u}_{\textnormal{new}}$ as below.
\begin{IEEEeqnarray}{rCl} 
\mathbf{y}_{(k+1)n_1} = [\mathbf{y}_{k n_1} , \mathbf{y}_{\textnormal{new}}],
\mathbf{u}_{(k+1)n_1} = [\mathbf{u}_{k n_1} , \mathbf{u}_{\textnormal{new}}],
\end{IEEEeqnarray}
where $\mathbf{y}_{\textnormal{new}} = [y_{k n_1 +1}, y_{k n_1 +2},..., y_{k n_1+n_1}]$ and $\mathbf{u}_{\textnormal{new}} = [u_{k n_1 +1}, u_{k n_1 +2},..., u_{k n_1+n_1}]$. However, the first $k n_1$ components of $\mathbf{a}_{k n_1}$ and $\mathbf{a}_{(k+1)n_1}$ are different. We need to derive the relationship with them.

Recall that $\mathbf{a}_n = \mathbf{1}_n\mathbf{C}_{\mathbf{y'}_n}^{-1}\{\mathbf{1}_n\mathbf{C}_{\mathbf{y'}_n}^{-1}\mathbf{1}_n^{\top}\}^{-1}$ from \eqref{eq:a_n and b}. The inverse of covariance matrix  $\mathbf{C}_{\mathbf{y'}_n}^{-1}$  has the following two properties.

\textbf{property 1}:
The matrix $\mathbf{C}_{\mathbf{y'}_n}^{-1}$ is similar to a diagonal matrix. That is, the entries outside the main diagonal are significantly smaller than the diagonal entries.

This property results from the covariance of $y_i'$ and $y_j'$, $E[(y_i-m_i)(y_j-m_j)]$, is quite small when $|i-j|$ is large. Accordingly, the matrix $\mathbf{C}_{\mathbf{y'}_n}$ is similar to a diagonal matrix, and so is its inverse matrix $\mathbf{C}_{\mathbf{y'}_n}^{-1}$.

\textbf{property 2}:
The diagonal entries of matrix $\mathbf{C}_{\mathbf{y'}_n}^{-1}$ repeats by the period of $n_1$ except the first submatrix. That is, the matrix $\mathbf{C}_{\mathbf{y'}_n}^{-1}$ shows as below.
\begin{IEEEeqnarray}{rCl} 
\mathbf{C}_{\mathbf{y'}_{(k+1)n_1}}^{-1} = \begin{bmatrix}
       \mathbf{A}_{n_1 \times n_1} & \mathbf{0} & \mathbf{0} & \hdots & \mathbf{0} \\[0.3em]
       \mathbf{0} & \mathbf{B}_{n_1 \times n_1} & \mathbf{0} &   & \mathbf{0} \\[0.3em]
       \mathbf{0} & \mathbf{0} & \mathbf{B}_{n_1 \times n_1} &   & \vdots \\
       \vdots &  &  & \ddots & \mathbf{0}\\
       \mathbf{0} & \mathbf{0} & \hdots & \mathbf{0} & \mathbf{B}_{n_1 \times n_1} \\
     \end{bmatrix}
\nonumber
\end{IEEEeqnarray}
This repetition property results from $y_{i+(k-1)n_1}'$ is similar to $y_{i+kn_1}'$ for $i=1,2,...,n_1$ and $2 \leq k \leq K-1$ because they all affect by similar level of ISI effect. We assume the crossover over two or more symbols can be ignored. Under this assumption, the effect of ISI on the second symbol is the same as the third one, and so are the following symbols. On the other hand, the first symbol without ISI causes the exception.

The above two properties of $\mathbf{C}_{\mathbf{y'}_n}^{-1}$ is useful when we find the relationship with $\mathbf{a}_{k n_1}$ and $\mathbf{a}_{(k+1)n_1}$. When $n = (k+1)n_1$, based on the properties of $\mathbf{C}_{\mathbf{y'}_n}^{-1}$, we can derive $\mathbf{a}_{(k+1)n_1}$ as 
\begin{IEEEeqnarray}{rCl} 
\label{eq:a_(k+1)n_1}
\mathbf{a}_{(k+1)n_1} 
&=& \frac{\mathbf{1}_n\mathbf{C}_{\mathbf{y'}_n}^{-1}}{\mathbf{1}_n \mathbf{C}_{\mathbf{y'}_n }^{-1}{\mathbf{1}_n}^{\top}}  
= \frac{\mathbf{1}_n\mathbf{C}_{\mathbf{y'}_n}^{-1}}{\mathbf{1}_{n_1}\mathbf{A}_{n_1\times n_1}\mathbf{1}_{n_1}^{\top} + k \mathbf{1}_{n_1}\mathbf{B}_{n_1\times n_1}\mathbf{1}_{n_1}^{\top}} \nonumber \\
&=& \frac{\mathbf{1}_{(k+1)n_1} \mathbf{C}_{\mathbf{y'}_{(k+1)n_1} }^{-1}}{A + k B},
\end{IEEEeqnarray}
where $A = \mathbf{1}_{n_1}\mathbf{A}_{n_1\times n_1}\mathbf{1}_{n_1}^{\top}$ and $B = \mathbf{1}_{n_1}\mathbf{B}_{n_1\times n_1}\mathbf{1}_{n_1}^{\top}$. Moreover, when $n=k n_1$, we can derive $\mathbf{a}_{k n_1}$ as
\begin{IEEEeqnarray}{rCl} 
\mathbf{a}_{k n_1} 
&=& \frac{\mathbf{1}_{k n_1}\mathbf{C}_{\mathbf{y'}_{k n_1}}^{-1}}{A + (k-1) B}.
\end{IEEEeqnarray}
In the same way, we split $\mathbf{a}_{(k+1)n_1}$ into tow parts, $\mathbf{a}_{\textnormal{old}}$ and $\mathbf{a}_{\textnormal{new}}$, where $\mathbf{a}_{\textnormal{old}}$ is the first $k n_1$ components of $\mathbf{a}_{(k+1)n_1}$ and $\mathbf{a}_{\textnormal{new}} = [a_{k n_1 +1}, a_{k n_1 +2},..., a_{k n_1+n_1}]$.
\begin{IEEEeqnarray}{rCl} 
\mathbf{a}_{(k+1)n_1} = [\mathbf{a}_{\textnormal{old}} , \mathbf{a}_{\textnormal{new}}]
\end{IEEEeqnarray}
In \eqref{eq:a_(k+1)n_1}, the denominator is a constant $A + k B$, and the nominator is a row vector which is the sum of all row vectors of $\mathbf{C}_{\mathbf{y'}_{(k+1)n_1} }^{-1}$. Accordingly, we can derive the relationship as below.
\begin{IEEEeqnarray}{rCl}
\mathbf{a}_{\textnormal{old}} 
&=& \frac{\mathbf{1}_{k n_1}\mathbf{C}_{\mathbf{y'}_{k n_1}}^{-1} + \mathbf{0}}{A+kB} 
= \frac{\mathbf{1}_{k n_1}\mathbf{C}_{\mathbf{y'}_{k n_1}}^{-1}}{A+(k-1)B} \frac{A+(k-1)B}{A+kB} \nonumber \\
&=& \mathbf{a}_{k n_1}\frac{A+(k-1)B}{A+kB}
\end{IEEEeqnarray}
\begin{IEEEeqnarray}{rCl}
\mathbf{a}_{\textnormal{new}} 
&=& \frac{\mathbf{0} +...+ \mathbf{0} + \mathbf{1}_{n_1}\mathbf{B}_{n_1 \times n_1}}{A+kB} 
= \frac{\mathbf{1}_{n_1}\mathbf{B}_{n_1 \times n_1}}{B} \frac{B}{A+kB} \nonumber \\
&=& \mathbf{w}_{n_1}\frac{B}{A+kB},
\end{IEEEeqnarray}
where $\mathbf{w}_{n_1}=\frac{\mathbf{1}_{n_1}\mathbf{B}_{n_1 \times n_1}}{B}$.

As a result, the estimator of ULE for $n = (k+1)n_1$ can be derived from the estimator of ULE for $n = k n_1$ by the iterative form as below.
\begin{IEEEeqnarray}{rCl} 
\hat{\tau}_{(k+1)n_1}
&:=& \mathbf{a}_{(k+1)n_1}(\mathbf{y}_{(k+1)n_1}-\mathbf{u}_{(k+1)n_1})^{\top} \nonumber \\
&=& \mathbf{a}_{\textnormal{old}}(\mathbf{y}_{k n_1}-\mathbf{u}_{k n_1})^{\top} + \mathbf{a}_{\textnormal{new}}(\mathbf{y}_{\textnormal{new}}-\mathbf{u}_{\textnormal{new}})^{\top} \nonumber \\
&=& \frac{A+(k-1)B}{A+kB} \mathbf{a}_{k n_1}(\mathbf{y}_{k n_1}-\mathbf{u}_{k n_1})^{\top} \nonumber \\
&+& \frac{B}{A+kB}\mathbf{w}_{n_1}(\mathbf{y}_{\textnormal{new}}-\mathbf{u}_{\textnormal{new}})^{\top} \nonumber \\
&=& \frac{A+(k-1)B}{A+kB} \hat{\tau}_{k n_1} 
+ \frac{B}{A+kB}\hat{\tau}_{\textnormal{new}},
\end{IEEEeqnarray}
where $\hat{\tau}_{\textnormal{new}} = \mathbf{w}_{n_1}(\mathbf{y}_{\textnormal{new}}-\mathbf{u}_{\textnormal{new}})^{\top}$.

For simplification, in IULE, we denote the previous estimator $\hat{\tau_{k}}$ as $\hat{\tau}_{k n_1}$ in ULE and the next estimator $\hat{\tau}_{k+1}$ as $\hat{\tau}_{(k+1)n_1}$ in ULE. Moreover, another parameter $\alpha = \frac{A}{B}$ represents the importance of the new estimator $\hat{\tau}_{\textnormal{new}}$ with respect to the previous estimator $\hat{\tau}_k$.
\begin{IEEEeqnarray}{rCl} 
\hat{\tau}_{k+1}
= \frac{\alpha +k-1}{\alpha +k} \hat{\tau}_{k} 
+ \frac{1}{\alpha +k}\hat{\tau}_{\textnormal{new}},
\end{IEEEeqnarray}
where $\alpha = \frac{A}{B}$.
In the case when $\alpha=1$, which means $A=B$, that is $T_s$ is large enough so that all symbols are almost independent with each other. Then, we treat $\hat{\tau}_{\textnormal{new}}$ with the same importance with $\hat{\tau}_k$ in this case. On the other hand, in the case when $\alpha > 1$, which means $A > B$, that is the following symbols are influenced by the ISI effects. Then, we reduce the importance of $\hat{\tau}_{\textnormal{new}}$ with respect to $\hat{\tau}_k$.

Moreover, because of the periodic property of $\mathbf{y}'$, $u_{i+(k-1)n_1} - u_{i+kn_1}$ is close to $T_s$ for $i=1,2,...,n_1$ and $2 \leq k \leq K-1$.  As a result, $\mathbf{u}_{\textnormal{new}} = k T_s \mathbf{1}_{n_1} + \mathbf{m}_{n_1}$ for some constant vector $\mathbf{m}_{n_1} = [m_1, m_2, ..., m_{n_1}]$.
\begin{IEEEeqnarray}{rCl} 
 \hat{\tau}_{\textnormal{new}} &=& \mathbf{w}_{n_1}(\mathbf{y}_{\textnormal{new}}-\mathbf{u}_{\textnormal{new}})^{\top} \nonumber \\
&=& \mathbf{w}_{n_1}(\mathbf{y}_{\textnormal{new}} - k T_s \mathbf{1}_{n_1} -\mathbf{m}_{n_1})^{\top}
\end{IEEEeqnarray}

To sum up, the algorithm of IULE is described as below. First, we compute $\mathbf{a}_{n_1}$ and $\mathbf{u}_{n_1}$ beforehand to initialize the first estimator $\hat{\tau}_1$. Second, we compute $\alpha$, $\mathbf{w}_{n_1}$, and $\mathbf{m}_{n_1}$ beforehand to iteratively update the previous estimator $\hat{\tau}_k$ for $1 \leq k \leq K-1$.
\begin{IEEEeqnarray}{rCl} 
\hat{\tau}_1 = \mathbf{a}_{n_1}(\mathbf{y}_{n_1}-\mathbf{u}_{n_1})^{\top}
\end{IEEEeqnarray}
\begin{IEEEeqnarray}{rCl} 
\hat{\tau}_{k+1}
= \frac{\alpha +k-1}{\alpha +k} \hat{\tau}_{k} 
+ \frac{1}{\alpha +k}\mathbf{w}_{n_1}(\mathbf{y}_{\textnormal{new}} - k T_s \mathbf{1}_{n_1} -\mathbf{m}_{n_1})^{\top}
\nonumber \\ 
\end{IEEEeqnarray}

\begin{table}[h]
    \centering
    \begin{tabular}{ll}
    \hline
    $\mathbf{a}_{n_1}$ & the weighted value of each arrival time without ISI \\
    &(derived from covariance matrix of the arrival time).\\\hline
    $\mathbf{u}_{n_1}$ & the mean vector of the arrival time without ISI. \\\hline
    $\mathbf{w}_{n_1}$ & the weighted value of each arrival time with ISI \\
    &(derived from covariance matrix of the arrival time).\\\hline
    $\mathbf{m}_{n_1}$ & the mean vector of the arrival time with ISI. \\\hline
    $\alpha$ & the weighted value between the previous estimation \\ 
    &and the new estimation.\\\hline
    \end{tabular}
    
\end{table}

The algorithm of IULE only needs the statistics of the first symbol without ISI, $\mathbf{a}_{n_1}$ and $\mathbf{u}_{n_1}$, the second or third symbol with ISI, $\mathbf{w}_{n_1}$ and $\mathbf{m}_{n_1}$, and the ratio of importance between them, $\alpha$. By these information, it is enough to iteratively derive ULE for large amount of molecules $n$ and large index of symbol $k$ without too much performance lost.

\subsection{Cramer-Rao Lower Bound}

According to the classic estimation theory, we analyze  the lower bound of variance for unbiased estimator, Cramer-Rao Lower Bound (CRLB). Here, we just consider the situation when $K=1$, so $n_1=N$ and $\mathbf{x}=\mathbf{0}_{n_1}$. From the example above, we know that $f(\mathbf{y'}|\mathbf{x}=\mathbf{0}_{n_1})$ can simplify to the order statistic of Inverse Gaussian distribution $f_T(t)$ in this special case. The following we derive the Fisher information number:
\begin{IEEEeqnarray}{rCl} 
I(n_1) &=& -E[\frac{\partial^2}{\partial \tau^2}\ln f(\mathbf{y}|\mathbf{x},\tau)] \nonumber \\
&=& -E[\frac{\partial^2}{\partial \tau^2}\ln \{{n_1}![\prod_{i=1}^{n_1} f_T(y_i-0-\tau)]\}] \nonumber \\
&=& -E[\displaystyle\sum_{i=1}^{n_1}
\{\frac{\partial^2}{\partial \tau^2}\ln f_T(y_i-\tau)\}] \nonumber \\
&=& -E[\displaystyle\sum_{i=1}^{n_1}
\{\frac{3}{2(y_i-\tau)^2}-\frac{\lambda}{(y_i-\tau)^3} \}] \nonumber \\
&=& -E[\displaystyle\sum_{i=1}^{n_1}
\{\frac{3}{2(y_i')^2}-\frac{\lambda}{(y_i')^3} \}] \nonumber \\
&=& -E[\displaystyle\sum_{i=1}^{n_1}
\{\frac{3}{2(T_{(i)})^2}-\frac{\lambda}{(T_{(i)})^3} \}] \nonumber \\
&=& -{n_1}E[\frac{3}{2(T)^2}-\frac{\lambda}{(T)^3} ]. 
\end{IEEEeqnarray}
In the last step of above derivation, because we treat $y_i'$ as order statistic of $T$, so we can rearrange them to treat summation result over $i$ as independent on the order of $T_{(i)}$. Accordingly, we can derive the fisher information number is the first order proportional to the quantity of molecules $n_1$. 
\begin{IEEEeqnarray}{rCl} 
Var[\hat{\tau}(\mathbf{y})] \geq \frac{1}{I(n_1)}
= \frac{1}{{n_1}E[\frac{\lambda}{(T)^3}-\frac{3}{2(T)^2}]}.
\end{IEEEeqnarray}
As a result, the CRLB is the first order inversely proportional to $n_1$.

\section{Blind Synchronization} \label{sec:blind}

In this section, we discuss the situation when $\{n_k\}$ for $1 \leq k \leq K$ is not constant but random for RN, because of the message embedded in the quantity level of molecules for every symbol. Therefore, different from above discussion, this case has no training phase anymore before communication. 

Following the paper \cite{lin_2012_signal}, we consider $M$-ary quantity-based modulation in general, that is $n_k \in \{L_0, L_1, ...,L_{M-1}\}$ as $M$ hypotheses. Accordingly, RN need to use the information in $\mathbf{y}$ to do both of synchronization (estimating $\tau$) and demodulation (detecting $n_k$).

\subsection{Non-decision-directed Parameter Estimation} 

Under the uncertainty of $\{n_k\}$, the releasing time sequence $\mathbf{x}$ is also random for RN. We can rewrite the joint pdf of observation $\mathbf{y'}$ when $\tau=0$ by averaging the conditional joint pdf over the probability of $\mathbf{x}$. Here, we assume \textit{a priori} probability of $\mathbf{x}$ is known for RN, so the joint pdf of $\mathbf{y'}$ can be derived as follows:
\begin{IEEEeqnarray}{rCl} 
f(\mathbf{y'}) = \displaystyle\sum_{ \mathbf{x}} \Pr\{\mathbf{x}\} f(\mathbf{y'}|\mathbf{x}). 
\end{IEEEeqnarray}

For simplification, we just apply ULE for $n=1$, that is $\hat{\tau}_{\textnormal{ULE}} := y_1 - E[y_1']$ ,which is a simple and efficient estimator. Obviously, we just use the first molecule in the first symbol duration to estimate $\tau$.
In general case when $T_s$ is large enough, the probability of crossover happen to all the molecules from the first symbol is quite small, so we almost can assume the first arrival molecule is from the first symbol. As a result, what we really care about is the uncertainty of $n_1$, because $y_1'$ is almost the first order statistic of $n_1$ iid random variables with generic distribution $f_T(t)$.
\begin{IEEEeqnarray}{rCl} 
f(y_1') &=& \displaystyle\sum_{l=L_0}^{L_{M-1}} \Pr\{n_1 = l\} f(y_1'|\mathbf{x}=\mathbf{0}_{l}) \nonumber \\
&=& \displaystyle\sum_{j=0}^{M-1} p_j f_{T_{(1)}^{(L_j)}}(y_1'),
\end{IEEEeqnarray}
where $T_{(i)}^{(l)}$ is the $i$-th order statistic of $l$ iid random variables with generic distribution $f_T(t)$ and $p_j$ is \textit{a priori} probability of $n_1 = L_j$. Accordingly, we can rewrite the ULE for $n=1$ as below:
\begin{IEEEeqnarray}{rCl} 
\hat{\tau}_{\textnormal{ULE}} := y_1 - \displaystyle\sum_{j=0}^{M-1} p_jE[T_{(1)}^{(L_j)}].
\end{IEEEeqnarray}
With this estimator, the theoretical MSE can be derived as below:
\begin{IEEEeqnarray}{rCl} 
E[{(\tau - \hat{\tau}_{\textnormal{ULE}})}^2] &=& Var[y_1']  
= \displaystyle\sum_{j=0}^{M-1} p_j\sigma_j^2 \nonumber \\
&+& \displaystyle\sum_{j=0}^{M-1}\displaystyle\sum_{i=0}^{j} (v_j-v_i)^2 p_i p_j,
\label{eq:th_ule_n_1}
\end{IEEEeqnarray}
where $\sigma_j^2=Var[T_{(1)}^{(L_j)}]$, and $v_j=E[T_{(1)}^{(L_j)}]$ for $j = 0,1,...,(M-1)$.

\subsection{Decision Feedback}

Because of blind synchronization, the output of demodulation is useful for timing synchronization. That is, the better the performance of detection is, the smaller the MSE of timing offset estimation will be, and vice versa. As a result, we can apply Decision Feedback (DF) method to improve ULE for $n=1$. 

The steps are listed as follows. First, get $\hat{\tau}_1 := y_1 - E[y_1']$ by ULE for $n=1$, and then demodulate the first symbol to get $\tilde{n}_1$ based on $\hat{\tau}_1$, where $\tilde{n}_1$ is the detected value of the quantity of the first symbol $n_1$. Finally, get $\hat{\tau}_{\textnormal{DF}} := y_1 - E[T_{(1)}^{(\tilde{n}_1)}]$ by DF method, where $E[T_{(1)}^{(\tilde{n}_1)}]$ is a derived random variable on the domain of $\{v_0, v_1, ...,v_{M-1}\}$.
%

To analyze the performance of DF, let $q_{i,j}$ denote the probability when RN detects $n_1$ as $L_j$ condition on $n_1 = L_i$, that is $q_{i,j} := \Pr\{\tilde{n}_1=L_j|n_1=L_i\}$. We can derive the theoretical MSE as below:
\begin{IEEEeqnarray}{rCl} 
E[{(\tau - \hat{\tau}_{\textnormal{DF}})}^2]
&=& \displaystyle\sum_{j=0}^{M-1} p_j\sigma_j^2 \nonumber \\
&+& \displaystyle\sum_{j=0}^{M-1}\displaystyle\sum_{i=0}^{j} (v_j-v_i)^2 (p_i q_{i,j} + p_j q_{j,i}).
\label{eq:th_df}
\end{IEEEeqnarray}

If we compare the \eqref{eq:th_df} with \eqref{eq:th_ule_n_1}, the only difference is the coefficient of $(v_j-v_i)^2$ term. Therefore, applying DF will improve performance when $p_i q_{i,j} + p_j q_{j,i} < p_i p_j$ for all $0 \leq i < j \leq M-1$, that is the average crossover error probability of level $i$ and $j$ is less than $p_i p_j$. Moreover, the best performance of DF can make MSE reach $\displaystyle\sum_{j=0}^{M-1} p_j\sigma_j^2$ when $q_{ij}=0$ for $i \neq j$, that is in the absence of demodulation errors, $\tilde{n}_1=n_1$, which known as Decision-directed Parameter Estimation.

\section{Numerical Results}
\label{sec:numerical}

In this section, we present the numerical MSE of three methods proposed in training-based synchronization, MLE and ULE, and two methods proposed in blind synchronization. Also, the theoretical curves and CRLB is shown as benchmark to evaluate the efficiency of estimators.

For the parameters of Additive Inverse Gaussian Channel in this paper, we set $T_a=298$($25^{\circ}$C), $\eta=8.9 \times 10^{-4}$(water in $25^{\circ}$C), $r=10^{-8}$($10$nm), $d=2 \times 10^{-5}$($20\mu$m), and $v=2 \times 10^{-6}$($2\mu$m/sec), so we use the random variable T$\sim$IG($\mu$,$\lambda$) with $\mu = 10$ and $\lambda = 8.1955$. 

\subsection{Training-based synchronization}

In this subsection, we present two simulation results in one case when $K=1$, which is just considering the first symbol, and another case when $K \geq 2$, which is considering multi-symbol with ISI effect.

\begin{figure}[!h]
\centering
\includegraphics[scale=0.25]{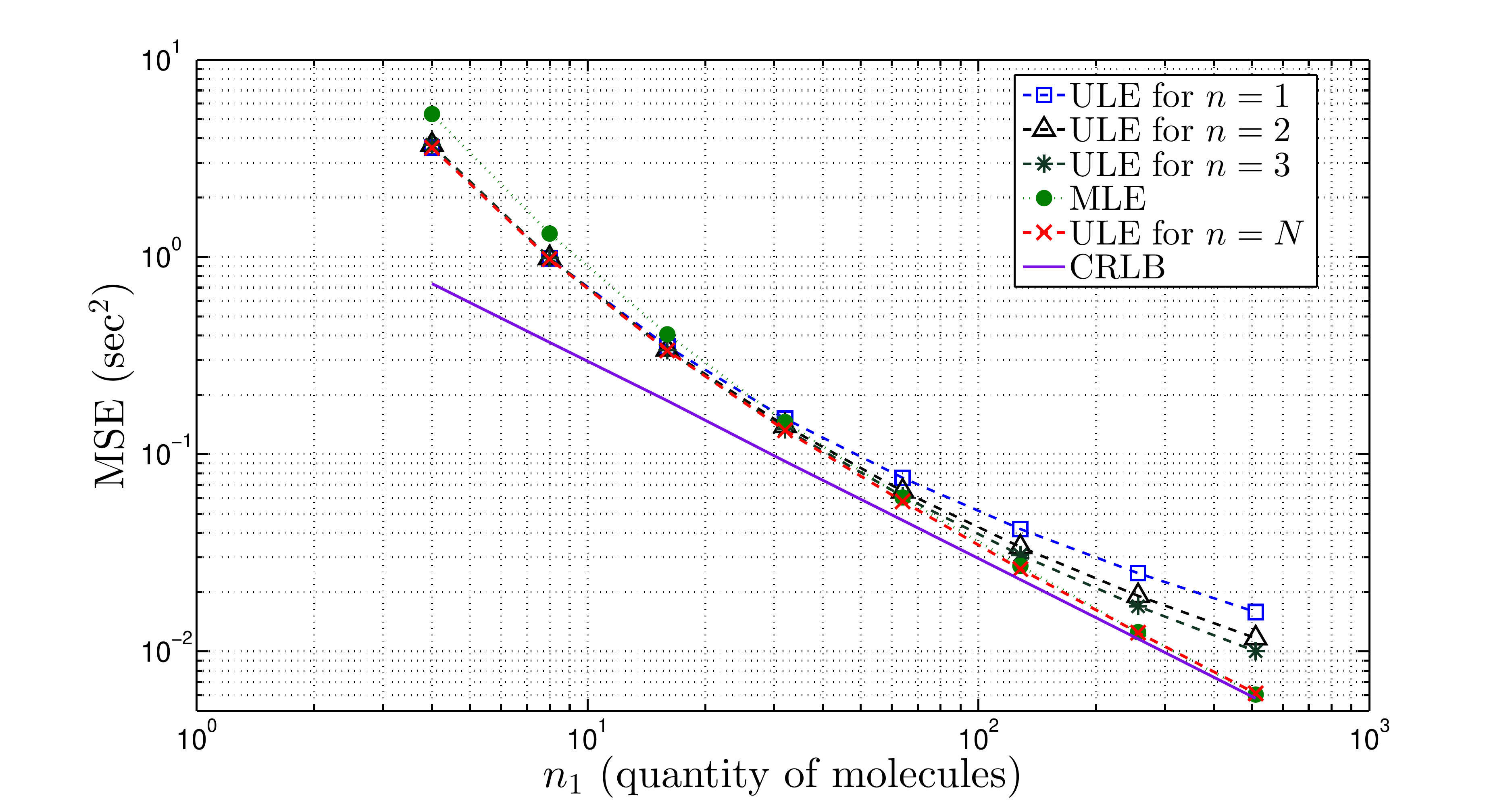}
\caption{The MSE of MLE and ULE with CMLB as benchmark in case when $K=1$, $n_1=N$, and 
$\mathbf{x}=\mathbf{0}_{n_1}$.}\label{ule_mleN_crlb}
\end{figure}

In Fig.~\ref{ule_mleN_crlb}, when we use ULE on $y_1, y_2, ..., y_N$, that is for $n=N$, to estimate $\tau$, the MSE can reach as small as using MLE, but the complexity of ULE reduce to linear time on $N$, which is quite lower than MLE. Considering the efficiency, we find out that the MSE of both two methods are close to CRLB when the quantity of molecules $N$ is large. Besides, ULE for just the first $n$ molecules can still reach a good estimation. It gives us an idea that the first $n$ arrival times include more useful information than the other observations.

The experiment in Fig.~\ref{ule_mleN_crlb} only simulates the special case when $\mathbf{x}$ is the zero vector. In general, when we use multi-symbol to estimate $\tau$, that is $K \geq 2$, the ISI effect will affect our result. In this case, we simulate ULE for $n=N$ when TN releases $K$ symbols with $n_1$ molecules per symbol, that is $n_1 = n_2 = ... = n_K$. Because ISI effect, the symbol duration $T_s$ affects the performance of estimation.
\begin{figure}[!h]
\centering
\includegraphics[scale=0.35]{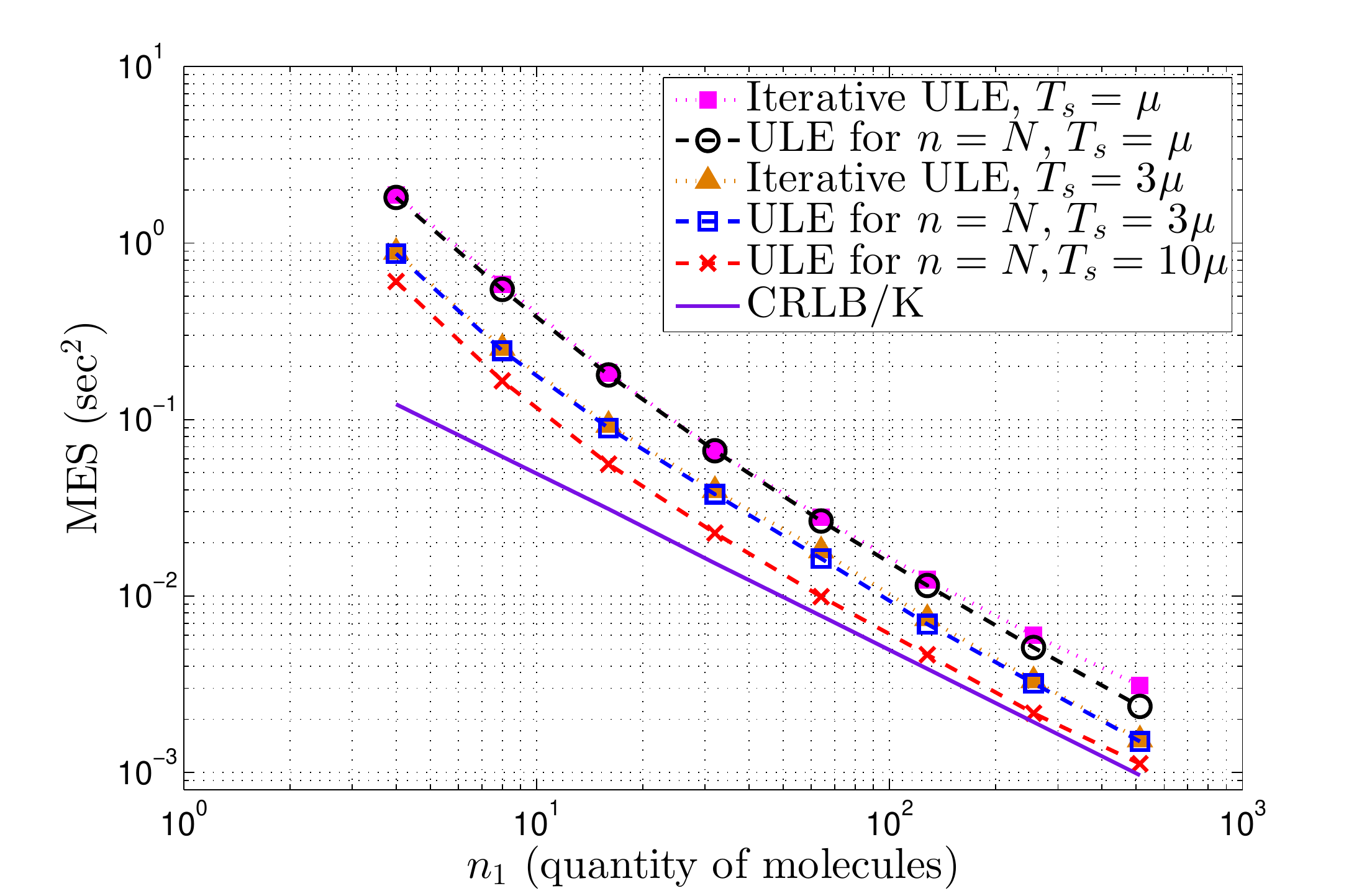}
\caption{The MSE of ULE for $n=N$ in case when $K=6$, and $n_1 = n_2 = ... = n_6$, where $\mu$ is the mean of $T$.}\label{ule_n_K_6}
\end{figure}

Fig.~\ref{ule_n_K_6} shows that MSE grows when $T_s$ is closed to $\mu:=E[T]$ due to ISI effect. On the other hand, when $T_s$ is large enough, all symbols become almost independent, so all symbols act as the first symbol without ISI effect. In this situation, we actually estimate $\tau$ with only one symbol and independently repeat this experiment $K$ times. Then, we average these $K$ estimation as the final result. Therefore, the MSE of ULE will approach to CRLB (the minimal variance of unbiased estimator for the first symbol)  divided by $K$, which matched in Fig.~\ref{ule_n_K_6}. Moreover, the MSE of Iterative ULE is really close to that of ULE for $n=N$, which verify that we do not lose too much information when we reform the iterative process in order to reduce the complexity. 


\subsection{Blind synchronization}

In blind synchronization, we present the numerical MSE of ULE for $n=1$, Decision Feedback, and Iterative Updating method. Also, we verify the result with theoretical curve and compare it with MSE of training-based methods as benchmark.

The parameters for $M$-ary quantity-based modulation is described as below; we set $L_j=\frac{(2j+1)n_1}{M}$ and $p_j = \frac{1}{M}$ so that $\sum_{j=0}^{M-1} p_j L_j = \bar{L} = n_1$ for $M=2,4,8$. Moreover, when $M=1$, the case reduce to training-based synchronization, which $n_1 = L_0$ is constant.
\begin{figure}[!h]
\centering
\includegraphics[scale=0.3]{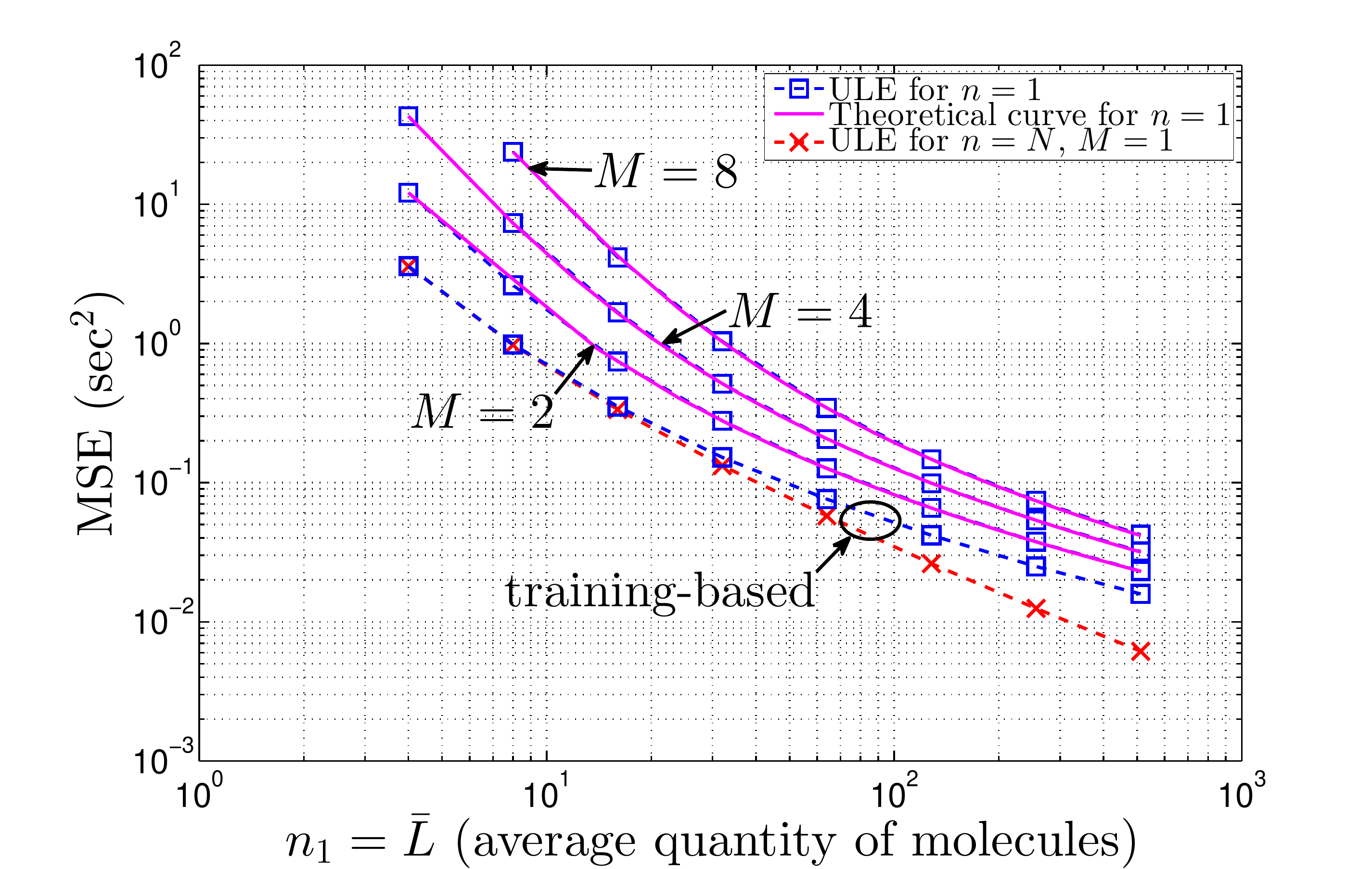}
\caption{MSE of ULE for $n$=1 and its theoretical curve in case when $K=1$, $T_s = 3\mu$, where $\mu$ is the mean of $T$.}\label{ule_1_M1248}
\end{figure}

Fig.~\ref{ule_1_M1248} shows that the performance of ULE for $n=1$ is worse when $M$ increases, which can be intuitively interpreted as the more random the message is, the harder we can estimate timing offset efficiently.
\begin{figure}[!h]
\centering
\includegraphics[scale=0.3]{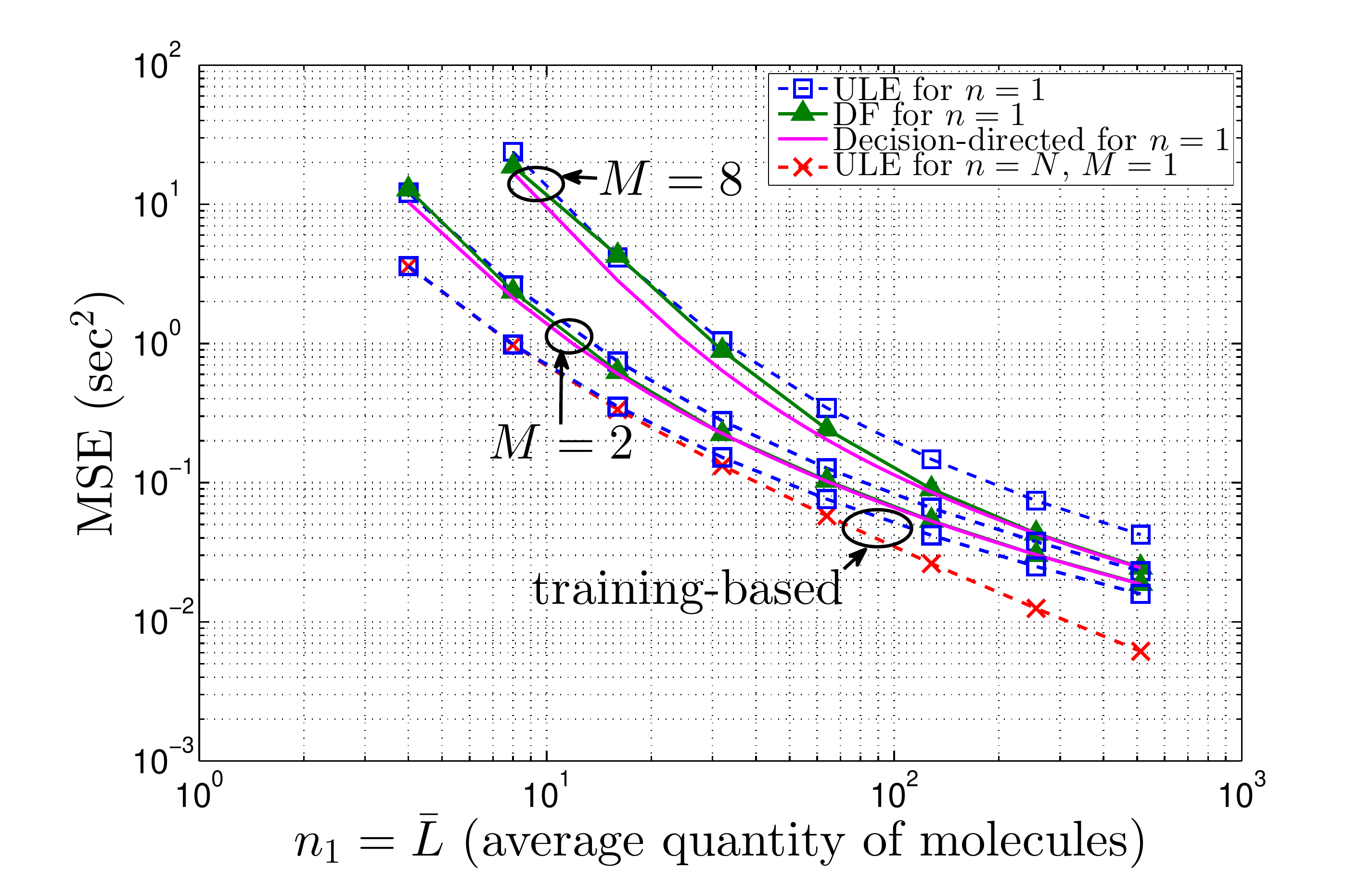}
\caption{MSE of Decision Feedback and theoretical Decision-directed Parameter Estimation in case when $K=1$, $T_s = 3\mu$.}\label{feedback_M28}
\end{figure}

In Fig.~\ref{feedback_M28}, we set the demodulation threshold at the middle of two adjacent quantity levels, that is $\tilde{n}_1:=\arg\min_{L_j}\left|L_j-l_1\right|$, where $l_1$ is the quantity of molecules in range of $(\hat{\tau}_1, \hat{\tau}_1 + T_s ]$. The numerical result shows that DF improves the MSE of ULE for $n=1$ and is closed to Decision-directed Parameter Estimation for this detection.

\section{Conclusion And Future Work} \label{sec:conclusion}

In this work, as far as we know, we first discuss the timing synchronization problem for quantity-based modulation in Additive Inverse Gaussian Channel. In training-based synchronization, we have proposed Iterative ULE, whose computational complexity is much lower than ULE and MLE. Moreover, its MSE reaches almost the same efficient level with ULE and MLE. On the other hand, we compare the theoretical MSE of ULE for $n=1$ with that of DF in blind synchronization, and give a sufficient condition when the latter improves the former. The next question we face is how accurate we need to estimate. To answer this question, the bit error rate has to be considered for future work. This analysis depends on the whole modulation and detection scheme we choose, which is more complicated and difficult to extent to general situation.


\bibliography{Ref_molecular}
\bibliographystyle{IEEEtran}
\end{document}